\documentclass[10pt]{article}

 \usepackage{amsmath,amsthm,amssymb}
 \usepackage{makeidx,epsfig,lscape}
 \usepackage{color,colortbl}
 \usepackage{fancyhdr}
 \usepackage{xcolor,pict2e}

 \renewcommand{\headrulewidth}{0pt}
 \renewcommand{\footrulewidth}{0.5pt}

 \definecolor{myaqua}{rgb}{0.0,0.5,0.55}
 \definecolor{lightaqua}{rgb}{0.75,0.95,0.95}

 \usepackage[colorlinks = true,
            linkcolor = myaqua,
            urlcolor  = blue,
            citecolor = myaqua]{hyperref}

\usepackage{caption}
\usepackage{floatrow}
\def\lin#1#2{\textcolor[rgb]{0.6,0.6,0.6}{\vspace*{#1mm} \hrule
   height 3 pt \vspace*{#2mm}}}
\def\bt{\begin{tabular}}
\def\et{\end{tabular}}
\def\and{\mbox{ and }}

\def\1{{\bf 1}}

 \def\sectionn#1{\refstepcounter{section}{\color{myaqua}

 \vskip 6mm

 \noindent\Large\bf\thesection. #1}

 \vskip 3mm}

 \def\boxx#1#2#3#4#5{
 {\linethickness{#4pt}\put(#1,#5){\color{myaqua}{\line(1,0){#3}}}}
 \multiput(#1,#2)(0,#4){2}{\line(1,0){#3}}
 \multiput(#1,#2)(#3,0){2}{\line(0,1){#4}}
  }

\begin{document}

 \fancyhead[L]{\hspace*{-13mm}
 \bt{l}{\bf International Journal of Astronomy \& Astrophysics, 2020, *,**}\\
 Published Online **** 2020 in SciRes.
 \href{http://www.scirp.org/journal/*****}{\color{blue}{\underline{\smash{http://www.scirp.org/journal/****}}}} \\
 \href{http://dx.doi.org/10.4236/****.2014.*****}{\color{blue}{\underline{\smash{http://dx.doi.org/10.4236/****.2014.*****}}}} \\
 \et}
 \fancyhead[R]{\includegraphics{pic1.ps}}

 $\mbox{ }$

 \vskip 12mm

{ % \fontfamily{Cambria}\selectfont

% "Title of the Paper"
{\noindent{\huge\bf\color{myaqua}
Near-Infrared Transit Photometry of Extra-Solar Planet HAT-P-54b
}}
%
% \runtitle{Change-Point Analysis of Survival Data}
\\[6mm]
{\large\bf Haruka Tabata$^1$ and Yoichi Itoh$^1$}}
\\[2mm]
{ %\fontfamily{Calibri}\selectfont
 $^1$ Nishi-Harima Astronomical Observatory, Center for Astronomy, 
University of Hyogo, 407-2 Nishigaichi, Sayo, Sayo, Hyogo 679-5313, Japan\\
Email: \href{mailto:yitoh@nhao.jp}{\color{blue}{\underline{\smash{yitoh@nhao.jp}}}}\\
 \\[4mm]
Received **** 2020
 \\[4mm]
Copyright \copyright \ 2020 by author(s) and Scientific Research Publishing Inc. \\
This work is licensed under the Creative Commons Attribution International License (CC BY). \\
\href{http://creativecommons.org/licenses/by/4.0/}{\color{blue}{\underline{\smash{http://creativecommons.org/licenses/by/4.0/}}}}\\
 \includegraphics{pic2.ps}

\lin{5}{7}

 { % \fontfamily{Cambria}\selectfont
 {\noindent{\large\bf\color{myaqua} Abstract}{\bf \\[3mm]
 \textup{
The results of near-infrared photometric observations
of a transit event of an extrasolar planet HAT-P-54b
are presented herein.
Precise near-infrared photometry was carried out using the
Nayuta 2 m telescope at Nishi-Harima Astronomical Observatory, Japan
and Nishi-harima Infrared Camera (NIC).
170 $J$-, $H$-, and $Ks$-band images
were taken in each band in 196 minutes.
The flux of the planetary system was observed to decrease during the transit event.
While the the $Ks$-band transit depth is similar to that in the $r$-band,
the $J$- and $H$-band transits are deeper
than those in the $Ks$-band.
We constructed simple models of the planetary atmosphere
and found that the observed transit depths are well reproduced
by inflated atmosphere containing H$_{\rm 2}$S molecule.
 }}}
 \\[4mm]
 {\noindent{\large\bf\color{myaqua} Keywords}{\bf \\[3mm]
planetary systems, infrared: planetary systems
}

 \fancyfoot[L]{{\noindent{\color{myaqua}{\bf How to cite this
 paper:}} Tabata et al. (2020)
Near-Infrared Transit Photometry of HAT-P-54b.
 International Journal of Astronomy \& Astrophysics,*,***-***}}

\lin{3}{1}

\sectionn{Introduction}

{ \fontfamily{times}\selectfont
 \noindent 
Astronomical observations of extrasolar planets have revealed a
rich diversity: for instance, planets
ranging from Earth-sized to giant Jupiter-sized ones;
besides, there are Jupiter-mass planets orbiting very close to the
host star, while there are others with an eccentric orbit.
Furthermore, the recent transit observations have revealed diversity 
not only on the planetary bodies and orbits, but also in terms of the
planetary atmosphere:
a fraction of the hot Jupiters have an inflating atmosphere, while
for others, the atmosphere seems to be escaping from the planet.
Detailed infrared observations of the transit events of
the extrasolar planets can be used to reveal the thermal structure and
temperature inversion of their atmosphere.

Photometric and spectroscopic observations of a transit event
reveal the chemical composition of the atmosphere
of an extrasolar planet.
Transit depth is deep at the wavelength where the atoms or molecules in
the atmosphere absorb the light of the host star.
To date, the presence of sodium, potassium \cite{Charbonneau}, 
and H$_{\rm 2}$O \cite{Deming} has been confirmed
in the atmosphere of the extrasolar planets,
while molecules such as CO, CO$_{\rm 2}$, and CH$_{\rm 4}$
have also been detected.
Furthermore, solid particles in the form of clouds and haze have also 
been identified
in the atmospheres of extrasolar planets.

HAT-P-54b is a hot-Jupiter orbiting a low-mass main-sequence star 
HAT-P-54 with an orbital period of 3.8 days. 
The mass and radius of the star are
0.645 solar-mass and 0.617 solar-radius, respectively.
The transit events of this planetary system were observed 
in the $r'$-band with the HAT-6 instrument at Fred Lawrence
Whipple Observatory (FLWO) in Arizona and the HAT-9 instrument
at Mauna Kea Observatory in Hawaii, and
in the $i$-band with the KeplerCam imager on the FLWO 1.2 m telescope \cite{Bakos}.
This observations yield that the planet-to-star radius ratio is 
0.157, i.e., the radius of the planet is 134,977 km, in the $r'$-band.
Combined with the Doppler shift measurements, the mass and orbital
inclination of the planet were derived to be 0.76 Jupiter mass and
87.04$^\circ$, respectively, using which, its the surface gravity was
deduced to be 21.09 m s$^{-2}$.

We conducted near-infrared photometry of the transit event
of HAT-P-54 using a unique three-color camera: the transit light curves
were simultaneously recorded in three bands.
Infrared wavelength observations offer a great advantage for investigating
the chemical composition of the atmosphere of an extrasolar planet,
because the surface temperature of most 
such planets is less than
2000 K, so that most species form molecules with vibration bands
in the infrared wavelengths.
}

\renewcommand{\headrulewidth}{0.5pt}
\renewcommand{\footrulewidth}{0pt}

 \pagestyle{fancy}
 \fancyfoot{}
 \fancyhead{} % clear all header and footer fields
 \fancyhf{}
 \fancyhead[RO]{\leavevmode \put(-90,0){\color{myaqua}H. Tabata, et al.} \boxx{15}{-10}{10}{50}{15} }
 \fancyhead[LE]{\leavevmode \put(0,0){\color{myaqua}H. Tabata, et al.}  \boxx{-45}{-10}{10}{50}{15} }
 \fancyfoot[C]{\leavevmode
 \put(0,0){\color{lightaqua}\circle*{34}}
 \put(0,0){\color{myaqua}\circle{34}}
 \put(-2.5,-3){\color{myaqua}\thepage}}

 \renewcommand{\headrule}{\hbox to\headwidth{\color{myaqua}\leaders\hrule height \headrulewidth\hfill}}

\sectionn{Observations and Data Reduction}

{ \fontfamily{times}\selectfont
 \noindent
The near-infrared transit observations of HAT-P-54 were
carried out on March 19, 2015
using the Nishi-harima Infrared Camera (NIC)
mounted on the Nayuta 2 m telescope at Nishi-Harima Astronomical Observatory,
Japan.
NIC has three 1024 $\times$ 1024 HgCdTe arrays with a field of view of 
2.73' $\times$ 2.73'.
The images in the $J$-, $H$-, and $Ks$-bands were taken simultaneously.
Near-infrared magnitudes of HAT-P-54 are 11.15 mag, 10.49 mag, and
10.33 mag in the $J$-, $H$-, and $Ks$-bands, respectively.
The transit event was predicted to begin at 21:48 and end at 
23:36 (JST).
Therefore, we observed the planetary system between 21:08 and 24:24 (JST).
As a reference star, 2MASS J06393662+2530129,
which is a main-sequence star located 1.37' away from HAT-P-54,
was also observed simultaneously.
Its near-infrared magnitudes are 10.95 mag, 10.37 mag, and 10.20 mag
in the $J$-, $H$-, and $Ks$-bands, respectively.
Ten frames of 30 s exposure each were taken with 10" radius telescope dithering.
In total, 170 frames were obtained in each $J$-, $H$-, and $Ks$-band.
The seeing condition was 3.0'', 2.7'', and 2.4'' in the $J$-, $H$-, and 
$Ks$-bands, respectively.

The data were processed using the Image Reduction and Analysis Facility
(IRAF). 
The object frames were calibrated in the standard manner, namely by dark
subtraction and flat fielding with the twilight frames.
Next, we created sky frames by calculating the median of the dithered 10 frames.
Each sky frame was then subtracted from the flat-fielded object frame.
A vertical stripe pattern remained in the sky-subtracted image.
We estimated the intensity of the stripe pattern in the region without objects,
and removed this count from the sky-subtracted image.
The fluxes of the target and reference stars were derived by aperture 
photometry:
the aperture radius was set to the FWHM of the PSF, so that the standard
deviation of the relative flux, as shall be explained later, 
shows minimum at the time out of the transit event.
The flux of the target star was divided by that of the reference star;
we call this the relative flux.
The relative fluxes before and after the transit should be the same,
unless the target or reference star is a variable star.
However, the average of 32 relative fluxes before the transit was
larger than that of 41 relative fluxes after the transit.
We fitted these 73 fluxes a linear function, and all relative 
fluxes were divided by
the function, so that the average of the out-of-transit relative flux
is unity.
Nevertheless, the relative flux still has an artificial pattern.
The position of the target and reference stars in the image were
changed by telescope dithering.
We noticed large relative flux 
when the target star was imaged in the lower part of the image,
which could be due to the imperfection of the flat fielding.
We calculated the average of the normalized relative fluxes out of the transit
for each dithering position, and then subtracted the average from each
normalized relative flux.
Finally, we added unity.
}

\sectionn{Results}
{ \fontfamily{times}\selectfont
 \noindent
The transit event of HAT-P-54b was detected in three bands in the near-infrared
wavelengths (Figure \ref{fig:lightcurve}).
We fitted the light curves to the relative flux at the EXOFAST website
(http://astroutils.astronomy.ohio-state.edu/exofast/exofast.html, 
\cite{Eastman}),
and then derived the transit depths.
The values listed in Table \ref{tab:hoststar} were used as the input
physical and orbital parameters for the host star and planet, respectively.
The transit depths are derived to be $0.0273\pm0.0010$, $0.0282\pm0.0007$,
and $0.0246\pm0.0010$ in the $J$-, $H$-, and $Ks$-bands, respectively.
The transit depths in the $J$- and $H$-bands are deeper than those in 
the $Ks$-band.
The EXOFAST website also provides the radius ratio of an extrasolar 
planet and its host star;
the derived radius ratios are larger
in the $J$- and $H$-bands than in the $Ks$-band 
(Table \ref{tab:radiusratio}).
Furthermore, we fitted a light curve to the $i$-band transit data 
obtained using
KeplerCam on the FLWO 1.2 m telescope
at JD 2456310 \cite{Bakos}.

\begin{figure}
  \begin{center}
   \begin{picture}(150,240)(0,-120)
    \put(-81,-138){\includegraphics{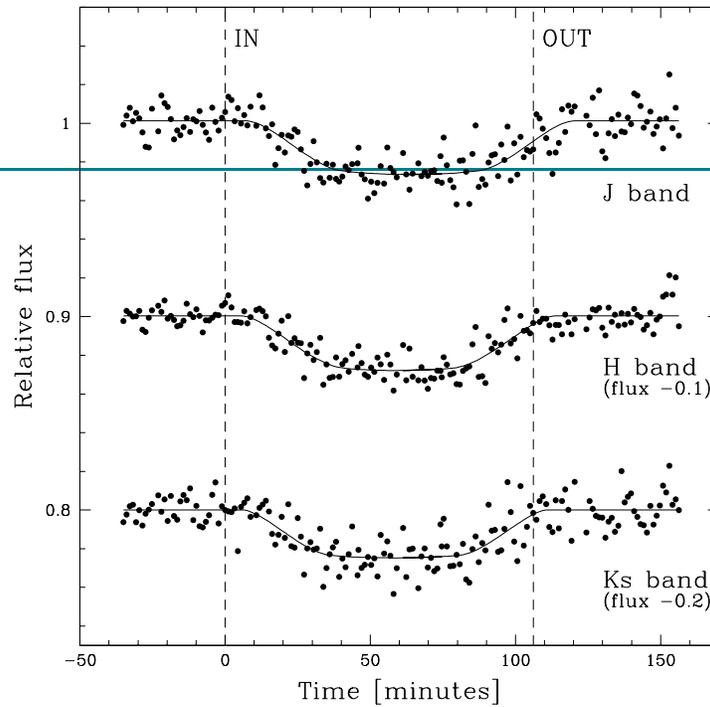}}
   \end{picture}
  \end{center}
  \caption{
Near-infrared relative fluxes of the transit event of HAT-P-54b.
The fluxes of HAT-P-54 were normalized by those of the reference star.
Solid lines represent the light curves
fitted with the differential evolution Markov chain Monte Carlo method.
        }\label{fig:lightcurve}
\end{figure}

\begin{table}
  \caption{Physical parameters of the host star and the orbital parameters of 
the planet}\label{tab:hoststar}
  \begin{center}
    \begin{tabular}{ll}
      \hline 
parameter & value \\
      \hline 
Effective temperature & $4390\pm50$ K \\
Metallicity & $-0.127\pm0.08$ \\
Surface gravity & $4.467\pm$0.012 [log cm s$^{-2}$] \\
Orbital period of the planet & $3.799847\pm1.4\times10^{-5}$ day \\
Orbital inclination of the planet & $87.040\pm0.084$ degree \\
      \hline 
    \end{tabular}
  \end{center}
\end{table}

\begin{table}
  \caption{Radius ratios of the planet to the host star}\label{tab:radiusratio}
  \begin{center}
    \begin{tabular}{ll}
      \hline 
band & radius ratio \\
      \hline 
$i$ & $0.162\pm0.001$ \\
$J$ & $0.165\pm0.003$ \\
$H$ & $0.168\pm0.002$ \\
$Ks$ & $0.157\pm0.003$ \\
      \hline 
    \end{tabular}
  \end{center}
\end{table}
}

\sectionn{Discussion}
{ \fontfamily{times}\selectfont
 \noindent

\begin{figure}
  \begin{center}
   \begin{picture}(150,240)(0,-120)
    \put(-81,-138){\includegraphics{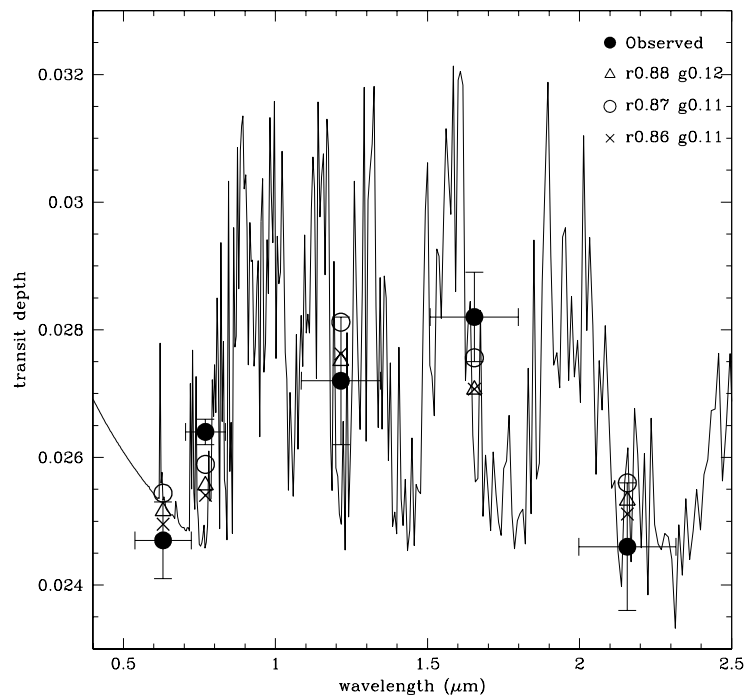}}
   \end{picture}
  \end{center}
  \caption{
The transit depths as a function of wavelengths.
The transit depth is the decrease of the
relative flux due to the planet's transit across the host star, normalized
by the relative flux out of the transit event.
Filled circles are the observed depths and the solid line shows the
transit depth spectrum of the model atmosphere.
Open circles indicate the transit depths of the
best-fit model atmosphere in optical and near-infrared
broad-bands, where the radius and surface gravity are
87 \% and 11 \% of the original values, respectively.
        }\label{fig:radiusratio}
\end{figure}

The transit depths are different in different bands.
The depth in the $Ks$-band is similar to the depth in 
the $r$-band \cite{Bakos}.
On the other hand, the depths in the $J$- and $H$-bands are
deeper than those in the $r$- and $Ks$-bands
(Figure \ref{fig:radiusratio}).

The transit depth does not change significantly
with wavelengths, if the planetary atmosphere
is cloudy or hazy.
However, if the planetary atmosphere is clear, the depth changes with 
the wavelength.
This is not only due to the absorption of photons from the host star
by the atoms and molecules in the atmosphere of the planet,
but also the Rayleigh scattering of the atoms and molecules at short wavelengths.
However, we cannot discuss the Rayleigh scattering for HAT-P-54b, owing to
the lack of transit observations at short wavelengths.

We constructed the transmission model spectra of the planet using the Planetary
Spectrum Generator \cite{Villanueva} (https://psg.gsfic.nasa.gov/), which
calculates the transit depths
as a function of the wavelengths.
The temperature of the host star was set to 4390 K,
its radius to 0.62 solar-radius, and 
the distance between the star and the planet to 0.0409 AU.
For the planetary atmosphere,
we used a pressure--temperature profile of a giant planet
\cite{Parmentier};
the equilibrium temperature was 818 K, and
no scattering aerosols were considered.
The free parameters were the radius of the planet, 
the scale height of the atmosphere,
and a species of an atom or molecule.
We varied the scale height by changing the value of the surface gravity.
We constructed atmospheric models by varying the radius and surface gravity
of the planet in steps of 1 \% of the original value, respectively.
The transit depths of the HAT-P-54 system 
have thus far been deduced only at five points by the photometric
observations.
To simplify the model,
only one atom or molecule species was included in the model atmosphere.
Sixty species of atoms or molecules were investigated.
We used the EXO-Transmit database \cite{Kempton} for
the cross section when available, and the HITRAN database otherwise.
Among the constructed transmission spectra, 
the spectra of the atmosphere containing OH, HCl, and H$_{\rm 2}$S
showed deep absorptions in the $J$- and $H$-bands.
We multiplied the transmission curves of the $r$-, $i$-, $J$-, $H$-, and $Ks$-band
filters to the transmission spectra and then calculated
the transit depth in each band.

The best fit spectrum was the spectrum of the atmosphere
containing H$_{2}$S molecules (Figure 2):
namely, the planetary radius of 117,430 km, 
which is 87 \% of the radius estimated from 
the $r'$-band transit data \cite{Bakos},
and
surface gravity of 2.32 m s$^{-2}$,
which is 11 \% of the surface gravity
calculated from the planetary radius derived from the $r'$-band
transit data and the planetary mass derived from the Doppler
shift measurement \cite{Bakos}
(Table \ref{tab:planet}).
The scale height of the atmosphere was about 10 times larger than
that in equilibrium, which indicates that 
the atmosphere of HAT-P-54b is inflating.
\begin{table}
  \caption{Physical parameters of the planet}\label{tab:planet}
  \begin{center}
    \begin{tabular}{ll}
      \hline 
parameter & value \\
      \hline 
Equilibrium temperature & 818 K \\
Radius & 117,430 km \\
Surface gravity & 2.32 m s$^{-2}$ \\
      \hline 
    \end{tabular}
  \end{center}
\end{table}

The sulfur chemistry in the atmosphere of hot Jupiters
investigated by several studies indicates
that the HS and S$_{\rm 2}$, which are
photochemically and thermochemically generated from H$_{\rm 2}$S, absorb
substantial amounts of light at high altitudes, causing stratospheric 
temperature inversions \cite{Zahnle}:
H$_{\rm 2}$S absorbs UV photons of wavelengths between 200 and
260 nm, and then dissociates into mercapto (HS) and H.
We estimated the UV-C flux ($\lambda$ = 121.6 -- 280 nm)
at the surface of HAT-P-54b.
The spectral type of the host star of the planet, i.e., HAT-P-54
is late-K.
The UV-C flux of the Sun at 1 AU is 3.38 Wm$^{-2}$ and that of a late-K
dwarf at 1 AU is 0.506 Wm$^{-2}$ \cite{Rugheimer}.
The semi-major axis of the orbit of HAT-P-54b is 0.0412 AU.
We calculated the UV-C flux at the surface of the planet as
$0.506 \times (\frac{1}{0.0412})^{2} = 298$ Wm$^{-2}$,
which is 88 times stronger than the flux at the surface of the Earth.

It was reported that while H$_{\rm 2}$S is abundant 
in the lower atmosphere ($P > 0.001$ bar),
CO is more abundant than H$_{\rm 2}$S
at any altitude \cite{Zahnle}, as it exhibits
strong absorption bands in the $Ks$-band.
However, our observations showed that the transit depth in the $Ks$-band 
is as shallow as that in the $r$-band;
thus, we did not find any evidence of CO absorption.

The observation possibility of 
H$_{\rm 2}$S in hot Jupiters using the JWST telescope is examined \cite{Wang}.
It is indicated that H$_{\rm 2}$S is the dominant species
at pressure levels between $1\times10^{-4}$ and $1\times10^{4}$ bar
in such atmospheres. 
The infrared model spectra of hot Jupiters observed by JWST were constructed in
the wavelength range from 2.0 to 11 $\mu$m:
The absorption features of H$_{\rm 2}$S between 2.6
and 2.8 $\mu$m
and between 3.5 and 4.1 $\mu$m can be detected
in the transmission spectra
for the planet with an equilibrium temperature higher than 1500 K.

We measured the transit depths of HAT-P-54b only in three bands,
and despite combining our results with those of a previous study, 
the transit depths could be determined
only in five bands by photometry.
Therefore, our conclusion that the atmosphere of HAT-P-54b contains H$_{\rm 2}$S may
not be a unique solution.
To seek other species and to confirm H$_{\rm 2}$S in the atmosphere of this
planet, photometric observations in other wavelengths and spectroscopic 
observations are required. 
More complicated atmospheric models including stratospheric
thermal inversion and altitude distribution of atoms and molecules are
mandatory for the comparison.
}

\sectionn{Conclusions}

{ \fontfamily{times}\selectfont
 \noindent
We conducted near-infrared photometry of the transit event of an 
extrasolar planet, HAT-P-54b.
Precise near-infrared photometry was carried out using the
Nayuta 2 m telescope
and Nishi-harima Infrared Camera (NIC).
170 near-infrared images
were taken in each $J$-, $H$-,
and $Ks$-band in 196 minutes.
The flux of the planetary system was observed to decrease during the transit event.
While the transit depth in the $Ks$-band was $0.0246\pm0.0010$,
which is similar to that in the $r$-band,
greater depths were observed in the $J$- and $H$-bands, i.e., $0.0273\pm0.0010$ and
$0.0282\pm0.0007$, respectively.
Furthermore, we constructed simple models of the planetary atmosphere
and found that the observed transit depths are well reproduced
by the inflated atmosphere containing the H$_{\rm 2}$S molecule.
}

 {\color{myaqua}

 \vskip 6mm

 \noindent\Large\bf Acknowledgments}

This work was supported by JSPS KAKENHI Grant Number JP17K05390.

 \vskip 3mm

{ \fontfamily{times}\selectfont
 \noindent

 {\color{myaqua}

}}

\end{document}